# A system built for both deterministic transfer processes and contact photolithography

Huandong Chen[1,4,*], Jayakanth Ravichandran[1,2,3]

[1]Mork Family Department of Chemical Engineering and Materials Science, University of Southern California, Los Angeles, CA, USA

[2]Ming Hsieh Department of Electrical and Computer Engineering, University of Southern California, Los Angeles, CA, USA

[3]Core Center for Excellence in Nano Imaging, University of Southern California, Los Angeles, CA, USA

[4]Present address: Condensed Matter Physics and Materials Science Department, Brookhaven National Laboratory, Upton, NY, USA

*Email: hchen3@bnl.gov

**Abstract**

A home-built compact system that functions as both a transfer stage for deterministic transfer processes and a mask aligner for contact photolithography, is constructed. The precision translation sample stage and optical microscope are shared between the two modes. In the transfer mode, assisted by either an adhesive or a heating element, the setup has been used to deterministically transfer freestanding semiconductors, 2D materials, and van der Waals electrodes. When configured in photolithography mode, the proposed instrument has been employed to fabricate various micro-scale patterns and devices, with minimum feature sizes of ~1-2 µm achieved. Our prototype instrument provides a feasible solution for preforming high quality deterministic transfer and photolithography processes on one single tool in house.

# Introduction

Deterministic transfer processes are essential for heterogeneously integrating different materials to achieve novel functionalities[1-3]. In recent years, due to the proliferation of the research on transfer printing of inorganic semiconductors for flexible electronics[4], CMOS-integration[5,6] and high-performance optoelectronic devices[7,8], as well as 2D materials and their heterostructures[9,10], a "transfer stage" becomes an essential instrument for fabricating such devices in many research labs in materials science, electrical engineering and physics. Basic components of a transfer stage include 1) linear substrate translation ($x$-, $y$- and $z$-axis) and rotation ($\theta$) stages for in-plane alignment, 2) a vertical sample translation stage ($z$-axis) for pickup and transfer, and 3) an optical microscopic system to assist these processes. Moreover, a heating stage is also commonly employed for thermally assisted transfer processes.

Besides, photolithography has been one of the bases for modern electronic, optoelectronic, and microfluidic device fabrication. In many research labs, photolithography processes are done in a shared central cleanroom facility using either a contact mask aligner or a maskless, direct writing system. Particularly, contact photolithography, which requires a photomask to be placed in contact or in proximity to a sample, plays a very important role in fundamental material and device research due to its simplicity and relatively high pattern resolution at micrometer scale, although such technique is no longer suitable for modern semiconductor manufacturing. A typical contact mask aligner consists of UV-source, photomask stage, optical microscope, translation and rotation sample stages.

One immediate thing to notice is that a contact mask aligner does share many of its major components, such as optical microscope, translation and rotation stages, with a transfer stage, for the purpose of precise in-plane alignment, despite their distinctive overall functionalities.

Therefore, naively thinking, by adding a UV-source and a photomask stage to a transfer stage or a second translation stage to an aligner, one could in principle integrate both functionalities onto one single setup. In this work, inspired by such a simple idea, we report the construction of such an instrument that works as both a high-precision transfer stage and a fully functional UV contact mask aligner. Our strategy provides a promising route for many research labs to perform both deterministic transfer and photolithography on one single compact system in house.

**Instrumentation**

The instrument was modified and upgraded from an existing rudimentary transfer stage that was initially built for picking up and transferring freestanding GaAs-based optoelectronic devices prepared from epitaxial lift-off processes (Figure S1). One major component of the original setup is a *Z*-translation stage (Series 443 linear stage, Newport) featuring a long travel distance (50 mm) that was installed vertically on a 90° angle bracket (360-90 bracket, Newport) to precisely adjust the sample-to-substrate distance. Besides, a stereotype microscope (3.5× - 90×, AmScope) was attached to an articulating arm to freely adjust the posture and a fiber-coupled halogen lamp (OSL2 fiber illuminator, Thorlabs) was used for illumination. All the components were installed on a honeycomb optical breadboard (IG-11-2, Newport) for the ease of assembly. In our initial design of the transfer stage, a stack of glass slide was used as the substrate stage that can be moved freely by hand for a rough alignment. Nonetheless, this transfer stage has been extensively used to fabricate various transfer-printed micro-scale GaAs-based solar cells and photoelectrochemical devices, with a majority of them not requiring high alignment accuracy[8,11-14].

We first upgraded the capability of precise in-plane translation alignment by installing two *XYZ*-translation stages (Series 562-XYZ, Newport) as the substrate stage and sample stage, respectively. An optional high-precision rotation stage (PR01, Thorlabs) can be attached to the substrate translation stage to achieve 5 arcmin (~ 0.08°) theta resolution, which is ideal for precise rotation alignment. Moreover, we implemented a simple heating stage to enable thermally assisted transfer processes, which has been widely adopted for fabricating various 2D material-based devices and their heterostructures. A small positive temperature coefficient (PTC) heating element (~ 20 × 25 mm, 50 W, self-regulating, Amazon) was attached onto a mounting plate using a double-sided foam tape (3M) to thermally isolate the heating element. A K-type thermocouple (Omega) was attached on the top surface of the heating element to monitor the temperature in real time and a benchtop DC power supply (E3610A, Agilent) was used to manually adjust the temperature. The PTC heating element could operate up to ~ 130°C with a supply voltage of ~ 5.5 V. This temperature range is ideal for various transfer processes using PDMS, polypropylene carbonate (PPC), and a large variety of thermally releasable tapes (Nitto "Revalpha" tape with 90°, 120°C release temperature, Semicorp.).

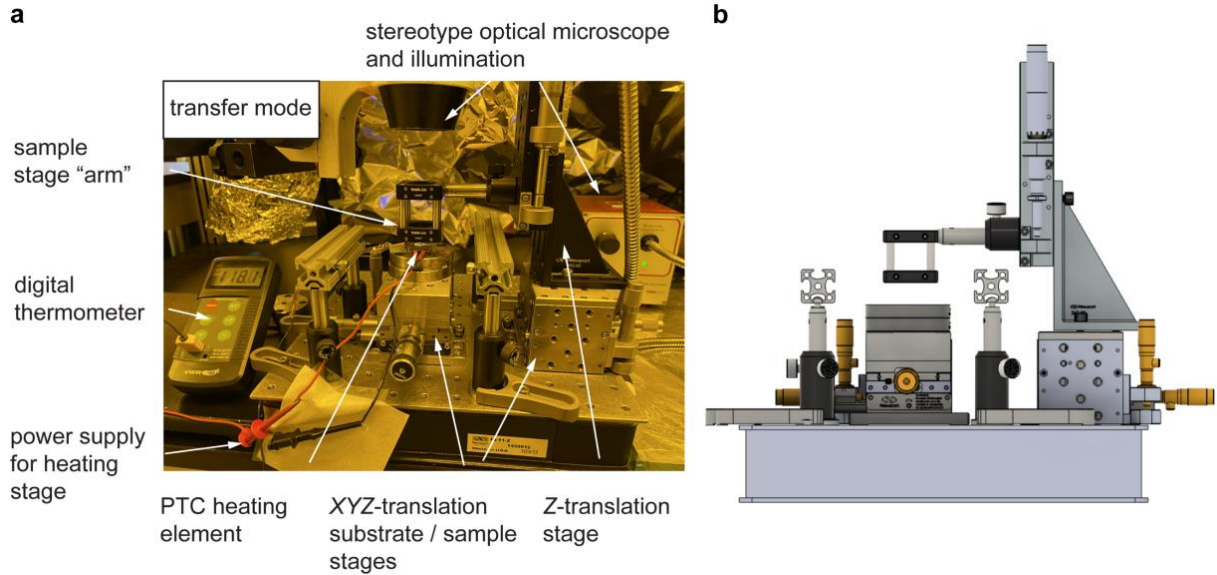

**Figure 1** (**a**) Optical image of the setup when configured to a transfer stage for thermally assisted transfer process. The configuration for adhesive-assisted transfer is similar except that the heating stage is not used. (**b**) Schematic illustration of the instrument in transfer mode (side view).

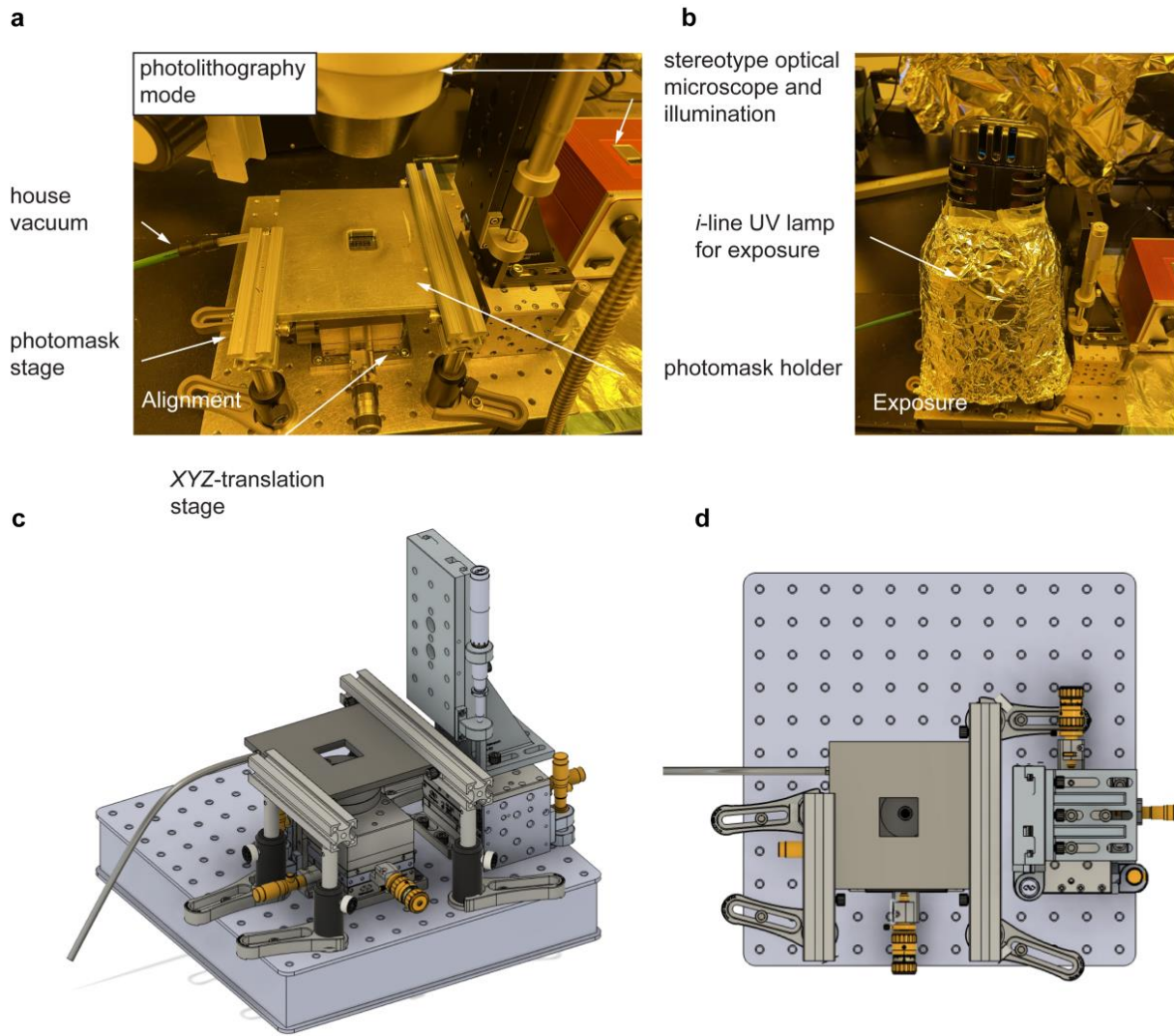

**Figure 2** (**a**) and (**b**) Optical image of the setup when configured into a contact mask aligner for photolithography during alignment and exposure procedures, respectively. A photomask stage and an *i*-line UV lamp is used in the photolithography mode in addition to the translation stages and optical systems that are shared between the two different configurations. (**c**) and (**d**) Perspective and top view of the schematic illustration of the instrument in photolithography mode.

Further, to convert the instrument to a working contact mask aligner for photolithography, a UV-source that photoresists are sensitive to, and a photomask holder are added. Here, we adopted a high intensity *i*-line UV lamp (UVP B-100AP, 100 W, 365 nm) as the light source, which was originally used for curing processes of UV-sensitive adhesives and epoxies in the lab. It is noted that such *i*-line light sources have been widely applied in contact photolithography processes to produce features down to several hundreds of nanometers in sizes till it hits the diffraction limit. One of the advantages of using such a large-area, high intensity lamp is to guarantee the uniform illumination on the samples, and hence, the photolithography results can be reliable and reproducible.

Lastly, we built a photomask holder stage surrounding the substrate translation stage, using sections of optical rails (25 mm square construction, Thorlabs) and optical post assemblies ($\Phi$1/2'', Thorlabs). The height of the posts was adjusted accordingly to accommodate Z-translation range of the substrate stage. A 3-inch photomask holder was machined (compatible to Karl Suss MJB3 mask aligner) and connected to the house vacuum line with an ON/OFF switch for photomask loading/unloading processes. Therefore, by simply adding a UV-source and a photomask stage, a transfer stage is successfully converted into a contact mask aligner. Regular 3-inch photomasks, either iron-oxide-based or chromium-based ones, can be used in the system.

**Deterministic transfer mode**

When the instrument is used as a transfer stage, the sample stage "arm", which consists of two optical cage plates (CP33T, Thorlabs) and several assembly rods ($\Phi$ 1/2'' and $\Phi$ 6 mm optical posts, Thorlabs), is first attached to the mounting board of the vertical Z-translation stage (Series 443 linear stage, Newport), while the photomask stage and UV-lamp are typically left unused. Removable double-sided tape (Scotch, 3M) was attached on the bottom cage plate to assist the sample stack (glass slide / PDMS / sample) mounting. Here, we present two examples of deterministic transfer processes, *i.e.*, adhesive-assisted transfer and thermally assisted transfer, using this setup:

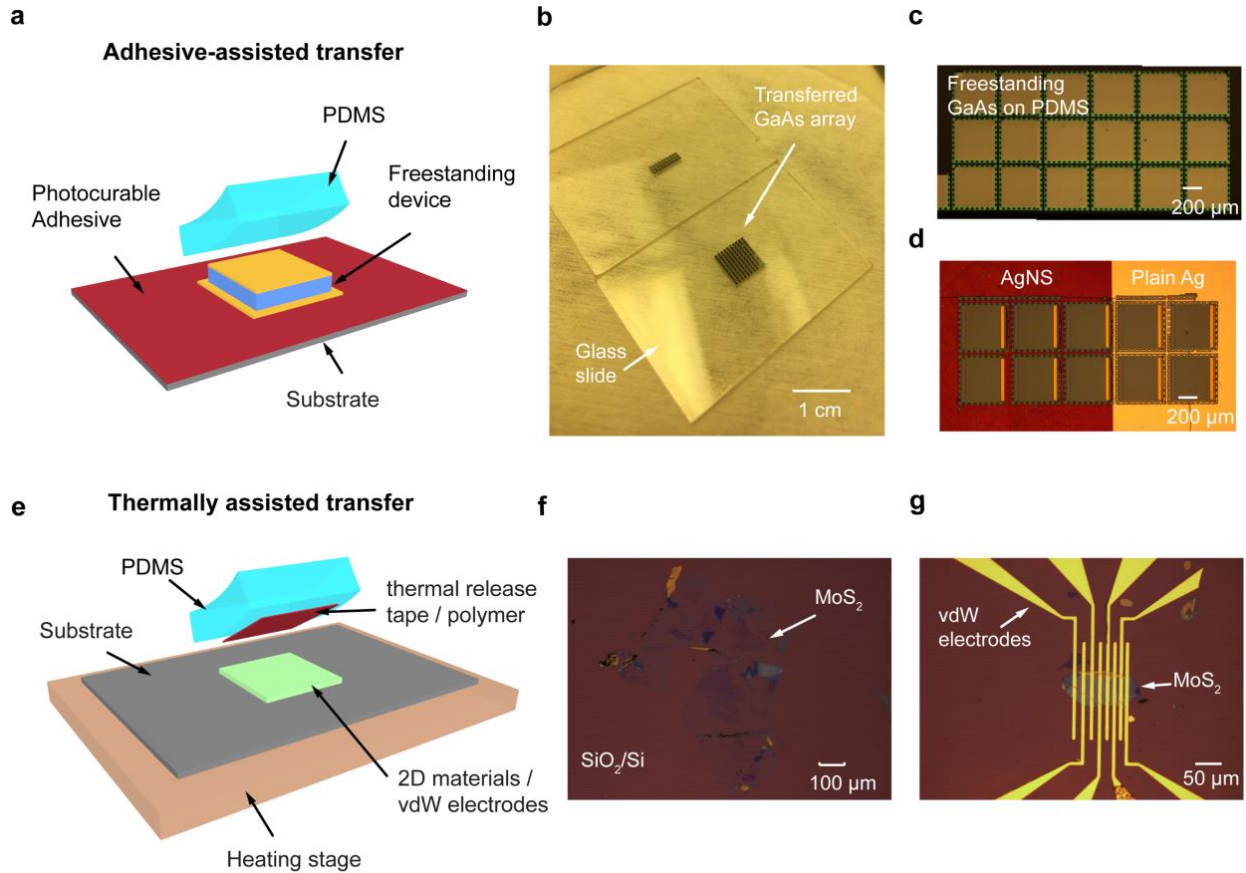

**Figure 3** (**a**) Schematic illustration of adhesive-assisted transfer process. (**b**) Optical image of a 10 × 10 micro-scale GaAs solar cell array transferred on a glass slide using a UV-curable adhesive. (**c**) to (**d**) Optical microscopic images of (**c**) freestanding GaAs devices picked up by a PDMS stamp after epitaxial lift-off, and (**d**) GaAs devices transferred on a plasmonic substrate with left 6 cell on nanostructured Ag substrate and the right 4 cells on plan Ag reflector. (**e**) Schematic illustration of thermally assisted transfer process. (**f**) to (**g**) Optical microscopic images of (**f**) few-layer MoS$_2$ transferred on a SiO$_2$/Si substrate, and (**g**) vdW electrodes integrated on a MoS$_2$ flake.

As illustrated in Figure 3a, adhesive-assisted transfer techniques have been widely used to facilitate the integration of freestanding semiconductor devices into desired functional layouts, e.g., back reflectors[13,15], flexible plastic sheets[16,17], glasses[12,14], etc., where sample-to-substrate interfaces are not critical in achieving the functionalities. In the scenario of transfer printing of micro-scaled GaAs photovoltaic devices, epitaxially grown device stacks were first fabricated into device arrays on wafer, and then released from the growth substrate *via* an epitaxial lift-off process, forming freestanding GaAs device arrays. A thin PDMS stamp (~ 5 mm × 5 mm × 1.5 mm) was attached to a glass slide and cut to form a sharp edge / corner using a razor blade before loading onto the sample stage. Selected arrays of GaAs devices were aligned to the stamp edge and then brought in contact with the PDMS stamp under the optical microscope. By separating the elastomer stamp from the source-wafer with a large releasing speed, the freestanding devices can be easily picked up due to a strong stamp-to-device adhesion. Figure 3c shows an optical microscopic image of the backside of a 3 × 6 GaAs device array picked up by a PDMS stamp.

We then used a thin (~ 1 µm) UV-curable spin-on-glass (SOG)-based adhesive to assist the transfer process. After brining the device arrays in contact with the adhesive, the PDMS stamp was slowly released to leave the GaAs devices "printed" on the new substrate. Figure 3b shows an optical image of a 10 × 10 GaAs device array transferred onto a glass slide to demonstrate the capability of large area transfer; Figure 3d shows an optical microscopic image of GaAs microcells transferred onto a plasmonic substrate, with six cells (left) sitting on a silver nanostructured reflector and the other four on a plain silver reflector. The detailed fabrication procedures of SOG adhesive-based transfer printing have been reported before[8,13,18], and part of the printed devices as shown in Figure 3b to 3d have been used to produce scientific results published elsewhere[13].

On the other hand, for heterogeneous integrations such as fabrication of 2D heterostructures[3,10] or dielectric gate integration[19,20], where the interfacial phenomena dominate, adhesive-free processes such as thermally assisted transfer are required, as illustrated in Figure 3e. The setup is fully compatible with standard dry transfer procedures for 2D materials, where a piece of transition metal dichalcogenide (TMDC) or graphite single crystal is usually mechanically exfoliated multiple times to achieve desired flake thickness, using a blue tape (Nitto) before loading onto a PDMS stamp, followed by pressing the whole stack onto a clean thermal oxide substrate using the transfer stage while turning on the power supply for the heating element (~ 5 V DC). The stamp can be slowly released when the substrate heats to ~ 90˚C, leaving 2D flakes transferred on the substrate with a high yield, due to an improved material-to-substrate adhesion. Figure 3f shows an optical microscopic image of few-layer $MoS_2$ transferred on a $SiO_2$/Si substrate following a gold-assisted transfer process for TMDC[21], and the color contrast indicates different number of layers.

In many scenarios, thermally releasable tapes or polymer layers, whose adhesive strength greatly reduces when heated beyond certain temperatures, are applied to improve the overall transfer yield. Figure 3g shows an example of integrated vdW electrodes on an exfoliated $MoS_2$ flake using PPC as the thermal-release layer. The vdW electrodes, or transferred electrodes, have been proven effective in reducing contact resistances to various semiconductors such as TMDCs[22,23] and halide perovskites[24,25], due to the elimination of damaged metal-to-semiconductor interfaces during metal deposition processes, and hence, improved contact interfacial quality is obtained. Here, the electrodes with desired patterns (~ 50 nm thick Au) were first fabricated on silicon wafers using regular photolithography-based microfabrication procedures. A photoresist layer (AZ 1518, Microchemicals) and PPC layer were spin-coated consecutively as the mechanical support and thermal-release layer, respectively, after which the whole stack was picked-up by a PDMS stamp using this setup. The freestanding electrodes were then aligned and transferred on the pre-prepared $MoS_2$ flake with a high yield by melting the PPC layer at ~110˚C, followed by a photoresist removal step using acetone. The detailed vdW electrodes transfer process using a photoresist / PPC stack was developed and optimized by the author, and it has been used to integrate transferred electrodes on a top-surface-planarized $BaTiS_3$ bulk crystal as reported elsewhere[26]. Moreover, it is worth noting that all the photolithography, pick-up and transfer processes involved in this example were performed using this setup in house. The detailed photolithography procedures using this instrument is discussed in the next section.

**Contact photolithography mode**

Aside from being used as a regular high-precision transfer stage for various heterogeneous integration, this setup can also be configured to the contact photolithography mode by simply adding a photomask stage and a portable UV lamp. To use the instrument as a contact mask aligner, a regular 3-inch photomask is first attached to a standard mask holder (machined, compatible to regular mask aligners) using house vacuum, with the patterns (either iron oxide or chromium) facing the top, after which the entire holder is slid into the photomask stage (made up of optical rails) and fixed using a positioning screw. The sample (coated with photoresist) is then brought to the proximity of the photomask by adjusting the $Z$-knob and aligned to the desired patterns under the optical microscope using the high-precision translation and rotation sample stage. Once all the aligning procedures are done, the sample is brought in contact with the photomask by fine-tuning the $Z$-knob. A thin PDMS sheet is often inserted below the sample as the buffer, which ensures a firm and conformal contact. Next, a portable UV-lamp with fixed intensity is used for flood exposure, and the desired doses are achieved by controlling the exposure time. The sample is then developed in a tetramethylammonium hydroxide-based developer (AZ MIF 726, MicroChemicals) to reveal the patterns following standard developing recipes. It is important to note that the operation procedures discussed above are similar to those of many manual contact mask aligners such as "Karl Suss MJB3", which are still being widely used in many universities' central cleanrooms for fundamental materials and devices research.

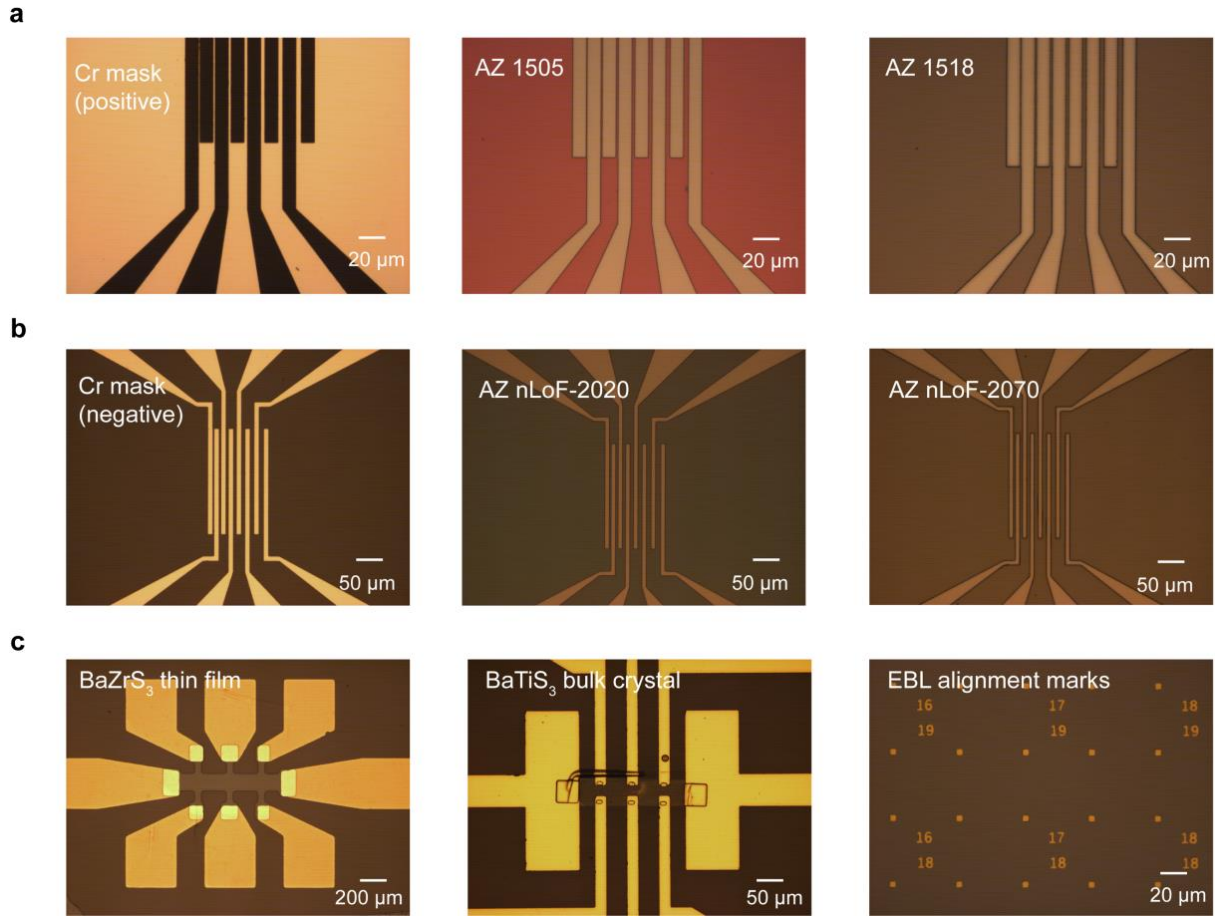

**Figure 4 Photolithography performance of the setup.** (**a**) Optical microscopic images of a TLM pattern (1 µm to 4 µm gap, 10 µm electrode width) on Cr photomask (left) and fabricated on Si substrates using positive photoresists: AZ 1505 (middle) and AZ 1518 (right). Feature sizes down to ~ 1 µm are clearly resolved with thin photoresist AZ 1505. (**b**) Optical microscopic images of a TLM pattern (2 µm to 10 µm gap, 10 µm electrode width) on Cr photomask (left) and fabricated on Si substrates using negative photoresists: AZ nLoF 2020 (middle) and AZ nLoF 2070 (right). (**c**) Optical micrograph images of a variety of devices fabricated using this setup as a contact mask aligner: a Hall bar device of $BaZrS_3$ thin film (left), a $BaTiS_3$ bulk crystal device used for transport measurements (middle), and alignment marks of 4 µm square arrays fabricated for Ebeam lithography (right).

Several commonly used photoresists, ranging from thick to thin, positive and negative, have been tested for the contact photolithography processes using this setup. Figure 4a and 4b show optical microscopic images of multi-electrode transmission line measurement (TLM) patterns using various photoresists such as AZ 1505 (positive, ~ 500 nm thick), AZ 1518 (positive,

~ 1.8 µm thick), AZ nLoF 2020 (negative, ~ 2 µm), and AZ nLoF 2070 (negative, ~ 7 µm), respectively. The exposure recipes for each photoresist were employed following optimized conditions that have been used by the author on other commercialized mask aligners, and no discernable difference was observed between these instruments. For example, feature sizes down to ~ 5 µm can be obtained using relatively thick photoresist such as AZ nLoF 2070, beyond which the photolithography resolution is limited by the photoresist film thickness. By combining a high-resolution chromium photomask and a thin photoresist (AZ 1505), feature sizes down to ~ 1-2 µm are achieved (Figure S3). Therefore, despite the simplicity of this mask aligner setup, we believe that its ultimate feature resolution shall be comparable to those commercialized aligners using the same *i*-line UV-source. The actual resolution achieved by a user is affected by many other factors such as photomask selection, conformal contact, photoresist thickness, exposure and developing conditions as well.

Figure 4c illustrates optical microscopic images of several actual devices fabricated using this setup for photolithography processes, including a "finger-shaped" photoconductive device based on $BaZrS_3$ thin film grown by pulsed laser deposition[27], a multi-electrode single-crystal $BaTiS_3$ device prepared for electrical transport measurements[28,29], and a set of aligning mark patterns prepared for Ebeam lithography (EBL) fabrication. Photoresist was selected based on the specific requirements of each application. For example, thin photoresist AZ 1505 is preferred for preparing patterns with fine features and high-resolution requirements such as the EBL aligning masks, while thick and negative resist AZ nLoF 2070 is commonly employed for fabricating relatively thick metal electrodes (300 – 400 nm thick), for the ease of lift-off.

**Limitations and potential modifications**

Nevertheless, this contact lithography setup does have a limitation on compatible sample sizes up to ~1 inch due to the limited travel distances of the sample translation stages. Samples with lateral sizes of 5 mm or 1 cm work best for this setup. Moreover, the optical microscope used here is simply a stereotype microscope hanging on an articulating arm, which could lead to limited in-plane alignment accuracy due to certain issues such as mechanical instability and limited optical magnification.

Table S1 lists all the optical and mechanical components used for constructing this simple but unique instrument that can be used as both a deterministic transfer stage and a contact mask aligner. The total cost is about ~ $12,000 if the instrument were built from scratch, which is already no way close to the listed price of any commercialized contact mask aligner or transfer stage. Importantly, since the assembly of the instrument is fully customizable, it offers large flexibility on its modifications. For instance, an upgrade on the optical microscope system, using a set of objective lenses with adjustable magnifications or even integrating a digital camera at the eyepiece port as most of the commercialized transfer stages and mask aligner have done, would greatly improve the alignment performance of this instrument for both deterministic transfer mode and contact photolithography mode. Moreover, one may even invest to fully motorize the translation motions for the ease of operation.

Alternatively, a downgrade modification is also possible to achieve a low-cost version that maintains both the functionalities and most of the performance. It is noted that most of the cost of the current instrument goes to the construction of a high-precision transfer stage, as high-precision optomechanical components from vendors such as Newport and Thorlabs can still be quite costly. On the other hand, several inexpensive instrumentation solutions have already been reported on constructing transfer stages for 2D materials over the past few years, and the total cost can be brought down to ~$1000 to $2000 with reasonably good performance[30,31]. Therefore, such solution will be advantageous to research groups with a tight budget.

## Conclusion

In conclusion, we have successfully constructed a compact multi-functional instrument that can be used as both a transfer stage for deterministic transfer processes and a mask aligner for contact photolithography. In the transfer mode, pick-up, adhesive-assisted and thermally assisted transfer of freestanding semiconductors, 2D materials, and vdW electrodes are demonstrated; and when the setup is used as a contact mask aligner, high quality photolithography processes with minimum feature sizes down to 1-2 µm are realized. Our prototype instrument offers an exciting opportunity for carrying out both transfer and photolithography processes in house.

## Author declarations

### Author contributions

H.C. conceived the idea and constructed the instrument. H.C. carried out transfer and photolithography experiments. H.C. and J.R. wrote the manuscript.

### Acknowledgements

This work was supported by an ARO MURI program (W911NF-21-1-0327) and the USC Viterbi School of Engineering. The authors gratefully acknowledge Mythili Surendran, Nan Wang, and Boyang Zhao for providing materials for testing. The authors thank Donghai Zhu for the help on preparing Cr photomasks using facilities at John O'Brien Nanofabrication Laboratory at USC.

**Data availability**

The data are available from the corresponding author of the article on reasonable request.

**Conflict of interest**

The authors declare no competing financial interests.

# TOC

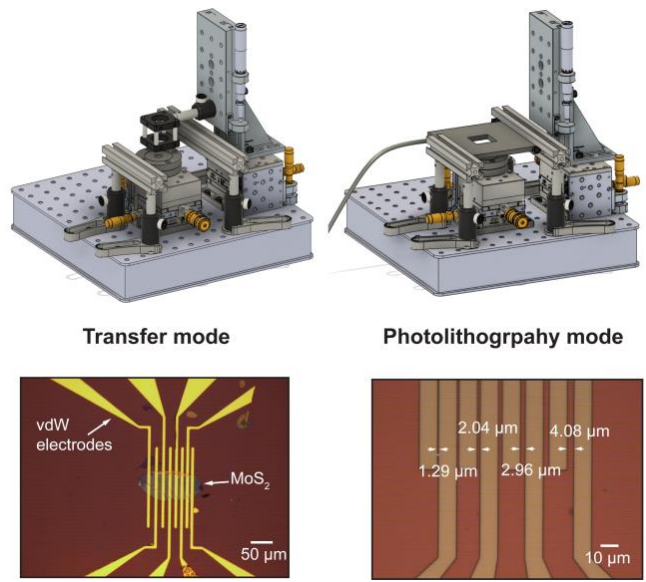

**Text:**

This work reports the construction of a compact system that functions as both a transfer stage for deterministic transfer processes and a mask aligner for contact photolithography. Our prototype instrument provides a feasible solution for performing both high quality transfer and photolithography processes on one single tool in house.

# Supplemental Information for

# A system built for both deterministic transfer processes and contact photolithography

**Authors:** Huandong Chen[1,4,*], Jayakanth Ravichandran[1,2,3]

[1]Mork Family Department of Chemical Engineering and Materials Science, University of Southern California, Los Angeles, CA, USA

[2]Ming Hsieh Department of Electrical and Computer Engineering, University of Southern California, Los Angeles, CA, USA

[3]Core Center for Excellence in Nano Imaging, University of Southern California, Los Angeles, CA, USA

[4]Present address: Condensed Matter Physics and Materials Science Department, Brookhaven National Laboratory, Upton, NY, USA

*Email: hchen3@bnl.gov

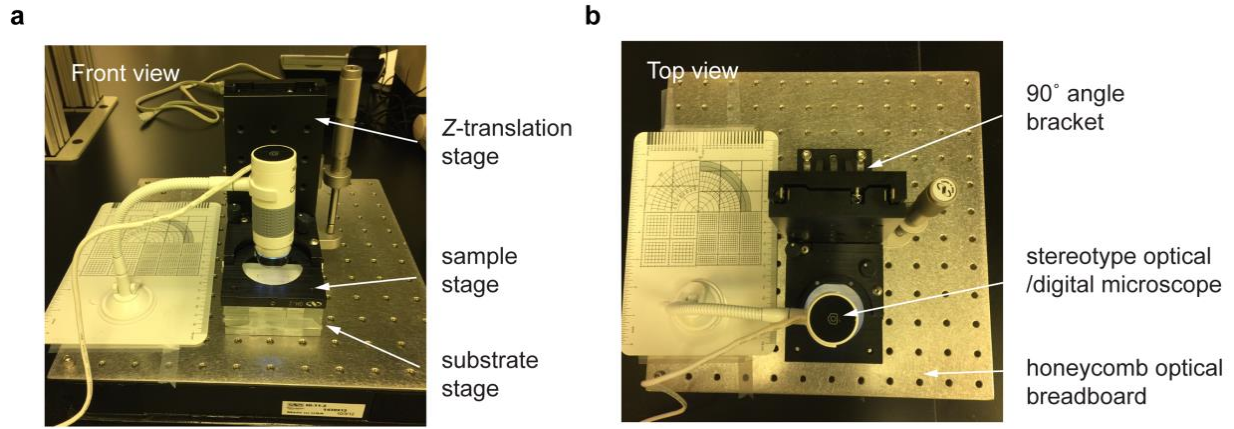

**Figure S1.** Optical image of the original transfer stage designed for transfer printing processes of freestanding semiconductors. (**a**) and (**b**) show the front view and top view of the setup, respectively. The original design uses a stack of glass slides as the substrate stage for manual alignment, and a portable digital microscope is used to take optical images during the alignment processes. The stereotype optical microscope is not included in the photo.

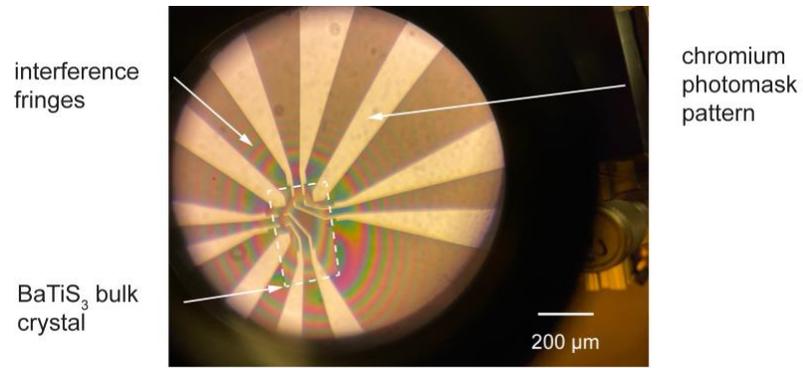

**Figure S2.** Optical microscope view of the photolithography alignment process of a planarized BaTiS$_3$ bulk crystal, observed through the eyepiece of the stereotype optical microscope. The dashed line indicates the size of the crystal, and the interference fringes indicate a conformal and firm contact between the chromium photomask pattern and the sample.

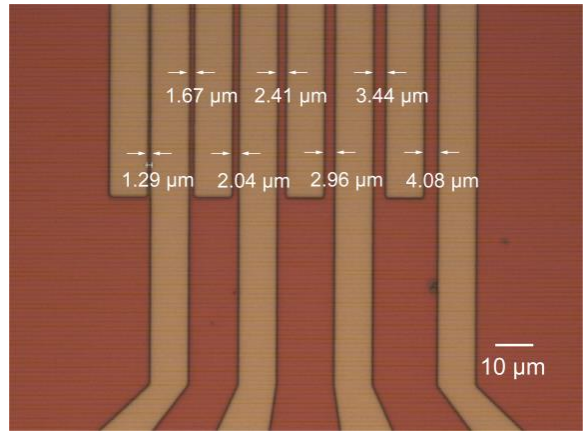

**Figure S3.** Optical microscopic image of a TLM pattern fabricated on Si substrate using AZ 1505 photoresist, with linewidth labeled on the image. The smallest line width resolved is ~ 1.29 µm.

| Components | Part number, vendor | Description | Cost | Purchase link |
|---|---|---|---|---|
| UV lamp | B-100 AP, UVP | 365 nm longwave UV, 100 watts intensity | $744 | https://www.fishersci.com/shop/products/uvp-blak-ray-b-100ap-high-intensity-uv-inspection-lamps-exposure-box/UVP76005501 |
| Optical microscope | SM-7 series, Amscope | 90 X stero zoom microscope on articulating stand with base plate | $824.99 | https://amscope.com/collections/stereo-microscopes/products/c-sm-7bz-48w |
| Illumination source for microscope | OSL2, Thorlabs | High-intensity fiber-coupled illuminator | $1086.94 | https://www.thorlabs.com/thorproduct.cfm?partnumber=OSL2 |
| Optical breadboard | IG-11-2, Newport | Honeycomb optical breadboard, 12 × 12 × 2.3 inch | $862 | https://www.newport.com/p/IG-11-2 |
| XYZ-translation stage | 562-XYZ, Newport | XYZ precision linear stage, with 12.7 mm travel distance | $2824, ×2 | https://www.newport.com/p/562-XYZ |
| Micrometer head | HR-13, Newport | Micrometer Head, High Resolution, 0.5 µm Sensitivity, 13 mm Travel | $222, ×4 | https://www.newport.com/p/HR-13 |
| Manual adjuster | DS-4F, Newport | Manual Adjuster, High Precision, 8.0 mm Coarse Travel, 0.3 mm Fine Travel | $437, ×2 | https://www.newport.com/p/DS-4F |
| Rotation stage | PR01, Thorlabs | High-precision rotation stage, 2.4 arcmin rotation per division | $386.07 | https://www.thorlabs.com/thorproduct.cfm?partnumber=PR01 |
| Z-translation stage | 443 series, Newport | High-performance linear stage, with 50 mm travel distance | $427 | https://www.newport.com/p/443 |
| Vernier micrometer | SM-50, Newport | Vernier Micrometer, 50 mm Travel, 23 lb. Load Capacity, 50.8 TPI | $314 | https://www.newport.com/p/SM-50 |
| 90° angel bracket | 360-90, Newport | Rigid 90° mounting plate with 1 inch separated slotted faces for mounting | $97 | https://www.newport.com/p/360-90 |
| Heating element | PTC heating element, Amazon | 2pcs PTC Heating Element 5W-50W | $11.99 | https://www.amazon.com/Bestol-Heating-consistant-Temperature-Thermostatic/dp/B07VBDT8NL |
| Sample mounting plate | PB1, Thorlabs | PB1 - Mounting Post Base, Ø2.48" x 0.40" Thick | $28.73 | https://www.thorlabs.com/thorproduct.cfm?partnumber=PB1 |
| Cage plate for sample mounting | CP35, Thorlabs | 30 mm cage plate with $\Phi$ 1'' double bore (CP02T was used for this used, but it has been discontinued) | $21.53 | https://www.thorlabs.com/thorproduct.cfm?partnumber=CP35 |
| Optical rails for constructing photomask stage | XE25RL2, Thorlabs | 25 mm square construction rail, unanodized | $46.04 | https://www.thorlabs.com/thorproduct.cfm?partnumber=XE25RL2 |
| Photomask positioning screw set | XE25T1, Thorlabs | Drop-in T-nuts for XE series rails | $34.17 | https://www.thorlabs.com/thorproduct.cfm?partnumber=XE25T1 |
| $\Phi$ 1/2'' optical posts | TR3, Thorlabs | $\Phi$ 1/2'' optical posts | $5.9 | https://www.thorlabs.com/thorproduct.cfm?partnumber=TR3 |
| Cage assembly rods | ER1.5, Thorlabs | Cage assembly rod, 1.5'' long, $\Phi$ 6 mm | $6.33 | https://www.thorlabs.com/thorproduct.cfm?partnumber=ER1.5 |

| Φ 1/2'' pedestal post holder | PH2E, Thorlabs | Φ 1/2'' pedestal post holder, L = 2.19'' | $28.51 | https://www.thorlabs.com/thorproduct.cfm?partnumber=PH2E |
| Clamping fork | CF175, Thorlabs | Clamping fork for Φ 1/2'' post holders | $11.76 | https://www.thorlabs.com/thorproduct.cfm?partnumber=CF175#ad-image-0 |

**Table S1** List of components required to build the setup from scratch.